\documentclass[twocolumn,prb,citeautoscript,superscriptaddress,8pt]{revtex4-2}

\usepackage{graphicx}
\usepackage{gensymb}
\usepackage{color}
\usepackage[dvipsnames]{xcolor}
\usepackage{float}
\usepackage{upgreek}
\usepackage{bm}
\usepackage{amsmath,amssymb}

\begin{document}

\title{Handheld device for non-contact thermometry via optically detected magnetic resonance of proximate diamond sensors}
%A handheld device for the contactless monitoring of optically detected magnetic resonances

\author{Gabriel J. Abrahams} % abrahams.gabi@gmail.com
\affiliation{School of Science, RMIT University, Melbourne, VIC 3001, Australia}

\author{Ethan Ellul} % s3661281@student.rmit.edu.au
\affiliation{School of Science, RMIT University, Melbourne, VIC 3001, Australia}

\author{Islay O. Robertson} % S3930288@student.rmit.edu.au
\affiliation{School of Science, RMIT University, Melbourne, VIC 3001, Australia}

\author{Asma Khalid} % asma.khalid@rmit.edu.au
\affiliation{School of Science, RMIT University, Melbourne, VIC 3001, Australia}

\author{Andrew D. Greentree} % andrew.greentree@rmit.edu.au
\affiliation{ARC Centre of Excellence for Nanoscale BioPhotonics, School of Science, RMIT University, Melbourne, VIC 3001, Australia}

\author{Brant C. Gibson} % brant.gibson@rmit.edu.au
\affiliation{ARC Centre of Excellence for Nanoscale BioPhotonics, School of Science, RMIT University, Melbourne, VIC 3001, Australia}

\author{Jean-Philippe Tetienne}
\email{jean-philippe.tetienne@rmit.edu.au}
\affiliation{School of Science, RMIT University, Melbourne, VIC 3001, Australia}

\begin{abstract} 

Optically detected magnetic resonance (ODMR) spectroscopy of defect-rich semiconductors is being increasingly exploited for realising a variety of practical quantum sensing devices. A prime example is the on-going development of compact magnetometers based on the nitrogen-vacancy (NV) defect in diamond for the remote sensing of magnetic signals with high accuracy and sensitivity. In these applications, the ODMR-active material is integrated into the overall apparatus to form a self-contained sensor. However, some emerging applications  require the sensing material to be in physical contact with an external object of interest, thus requiring an independent readout device. Here we present an ODMR-meter, a compact device specially designed to allow convenient, contactless monitoring of ODMR in a target object, and demonstrate its application to temperature monitoring with NV defects. Our prototype is composed of a handheld readout head (integrating all the necessary optical components and a microwave antenna) and a control box connected to a laptop computer, all made primarily from commercial off-the-shelf components. We test our device using an NV-rich bulk diamond as the object, demonstrate a temperature sensitivity of $10~{\rm mK}/\sqrt{\rm Hz}$ in static conditions, and demonstrate the feasibility of handheld operation. The limitations to measurement speed, sensitivity and accuracy are discussed. The presented device may find immediate use in medical and industrial applications where accurate thermometry is required, and can be extended to magnetic field measurements.       

\end{abstract}

\maketitle 

\section{Introduction}

Optically detected magnetic resonance (ODMR) is a spectroscopy technique that allows electron spin transitions in a substance to be probed via its optical response~\cite{Kohler1999}. This technique has gained popularity in the last two decades in the context of quantum science and technology, as it offers a way to study quantum systems in ambient conditions with a relatively simple experimental apparatus~\cite{Awschalom2018,Atature2018}. In this context, ODMR is typically associated with point defects in semiconductors, the most popular of which being the nitrogen-vacancy (NV) defect in diamond~\cite{Doherty2013}. An important application of ODMR is in precision sensing, which exploits the dependence of the spin resonances on physical quantities such as magnetic field, electric field, and temperature~\cite{Rondin2014,Schirhagl2014,Casola2018,Scholten2021}. Importantly, ODMR-based sensing can be highly accurate because the property being measured (spin resonance frequency) often has a well defined (and constant over time) relationship with the quantity of interest, e.g. magnetic field via the Zeeman effect~\cite{Degen2017,Barry2020}. Indeed, point defects such as the NV in diamond are atom-like systems with quantized energy levels that are robust against a wide range of perturbations~\cite{Doherty2013}, making it possible to use them as absolute, calibration-free sensors -- or ``quantum'' sensors.   
In practice, ODMR-based sensing requires four main ingredients~\cite{Bucher2019}: (i) an ODMR-active material, for instance a diamond containing NV defects; (ii) a light source to optically excite the material, typically a laser or light-emitting diode (LED); (iii) a microwave (MW) source to drive the spin transitions; and (iv) a light detection system to measure the spin-dependent optical response of the material (typically via its photoluminescence, less commonly by measuring absorption~\cite{Rondin2014}). Each application comes with its own set of requirements, placing specific constraints on the design of the apparatus combining the above four ingredients to realise a sensing device. For instance, applications in nano/microscale sensing are typically built around complex optical microscope setups in a laboratory setting to achieve high spatial resolution in the optical excitation and readout. The ODMR-active material is placed in the microscope in close proximity (nanometres to micrometers) to an object of interest, for instance to measure the localised stray magnetic fields emanating from a microscale object. The MW source generally consists of an antenna placed close to the sensor-object assembly. Example implementations using NV defects include confocal-based sensing, scanning probe microscopy, fiber-based sensing, and widefield imaging, which are increasingly used in condensed matter physics and other scientific applications~\cite{Rondin2014,Schirhagl2014,Casola2018,Scholten2021}. The overall apparatus is complex and inherently open since it must be possible to exchange the object under study. 

Another class of applications of ODMR-based sensing concerns remote sensing, where the source of the field to be measured originates from an extended or physically remote object (metres to kilometres or more), typically geomagnetic fields and their perturbations due to distant metallic objects, such that the field is uniform over the length scale of the sensing apparatus. In this case, the apparatus can be completely enclosed (as there is no object to exchange) and made very compact due to relaxed spatial resolution requirement for optical excitation and readout. In recent years, motivated by applications such as magnetic navigation and magnetic anomaly detection, intense efforts have been devoted to optimally integrate the four necessary ingredients to meet given size, weight and power requirements. In particular, compact and portable NV magnetometers have been demonstrated where the sensing head integrates most of the four ingredients in a small package~\cite{Zhang2017,Kim2019,Sturner2019,Webb2019,Zheng2020,Huang2021,Sturner2021,Mariani2022,Wang2022,Xu2022}. These devices are now being tested in real-world settings -- outside laboratories -- for various magnetometry applications.  

Between the two extremes of nanoscale sensing and remote sensing lies the sensing of macroscopic objects (millimetres to centimetres), for example for the magnetic detection of biological activity (e.g., magnetoencephalography or magnetocardiography) and the monitoring of electric currents and temperature in batteries~\cite{Barry2016,Webb2021,Arai2022,Hatano2022}. These applications are realised using a sensor head that integrates the ODMR-active material, typically a bulk diamond. The sensor head is then brought close to the target object, sometimes with physical contact. Fiber-based solutions that separate the optical components from the diamond have been developed to facilitate the sensor-object interfacing ~\cite{Kuwahata2020,Patel2020,Chatzidrosos2021,Hatano2021}. However, integrating the active material into the sensing apparatus is not always desirable. In particular, monitoring the temperature of an object is best achieved by physically contacting it to the ODMR-active material while keeping the rest of the measuring apparatus thermally decoupled, hence physically separate. This configuration has been explored in the context of nanoscale sensing, by incorporating diamond nanoparticles into or onto the object and reading them out using an optical microscope~\cite{Simpson2017,Andrich2018,Foy2020,Fujiwara2020,Khalid2020}, but extension to macroscopic objects has received little attention to date, and is the focus of the present work.

In particular, it has been recently proposed that biodegradable dressings incorporating diamond particles~\cite{Fox2016,Guarino2020} could be used for wound monitoring via temperature sensing~\cite{Khalid2020}.
More generally, we envision that such ``smart'' dressings or coatings could find use in a range of medical and industrial applications where accurate temperature measurements are required. For the proposed wound monitoring application, one needs to be able to measure temperatures in the biological window (25 to 40$^\circ$C) with a sub-kelvin accuracy and sensitivity to allow the reliable detection of wound infection, which can lead to a temperature increase by up to 5$^\circ$C at the wound site~\cite{Chanmugam2017,Houshyar2021}. Standard non-contact thermometry methods such as infrared thermometry have been considered, but their inaccuracy (caused by the dependence of infrared emission on various extraneous parameters) limits their reliability. ODMR-based thermometry of a smart wound dressing has the potential to meet the requirements and thus provide a new way to monitor infection of a wound without the need for dressing removal.

In these smart dressing/coating scenarios, the ODMR-active material (e.g. diamond particles) is physically attached to the target object and needs to be read out using a separate device at a distance, without any physical contact. The readout head should ideally be compact, integrate all necessary components (including a MW antenna), and be capable of handheld operation. The design requirements thus share similarities with the remote sensing applications discussed above, but one major difference is the fact that the MW antenna is integrated into the readout head, away from the active material. This poses additional constraints on the overall design and places a limit on the practical working distance achievable (i.e. the distance between readout head and object). In this work, we present a compact device specially designed to meet these various constraints and enable non-contact readout of macroscopic ODMR-active materials, i.e. an ODMR-meter.

Our ODMR-meter is composed of a handheld readout head (integrating the optical components and a MW antenna) and a control box connected to a laptop computer with a purpose-built software user interface. The presented prototype is made primarily from commercially available, off-the-shelf components and, while already relatively compact, can be further optimised through custom engineering to reduce size and weight. We test our device using a standard bulk diamond as the active material and demonstrate a temperature sensitivity of $10~{\rm mK}/\sqrt{\rm Hz}$ in static conditions. We also demonstrate the feasibility of handheld operation. The limitations to measurement speed, sensitivity and accuracy are discussed. The presented device may find immediate use in medical and industrial applications where accurate thermometry is required, and can be extended to magnetic field measurements. 

\section{Design of the ODMR-meter}

\subsection{Device concept}

\begin{figure*}[tb!]
    \centering
    \includegraphics[width=17cm]{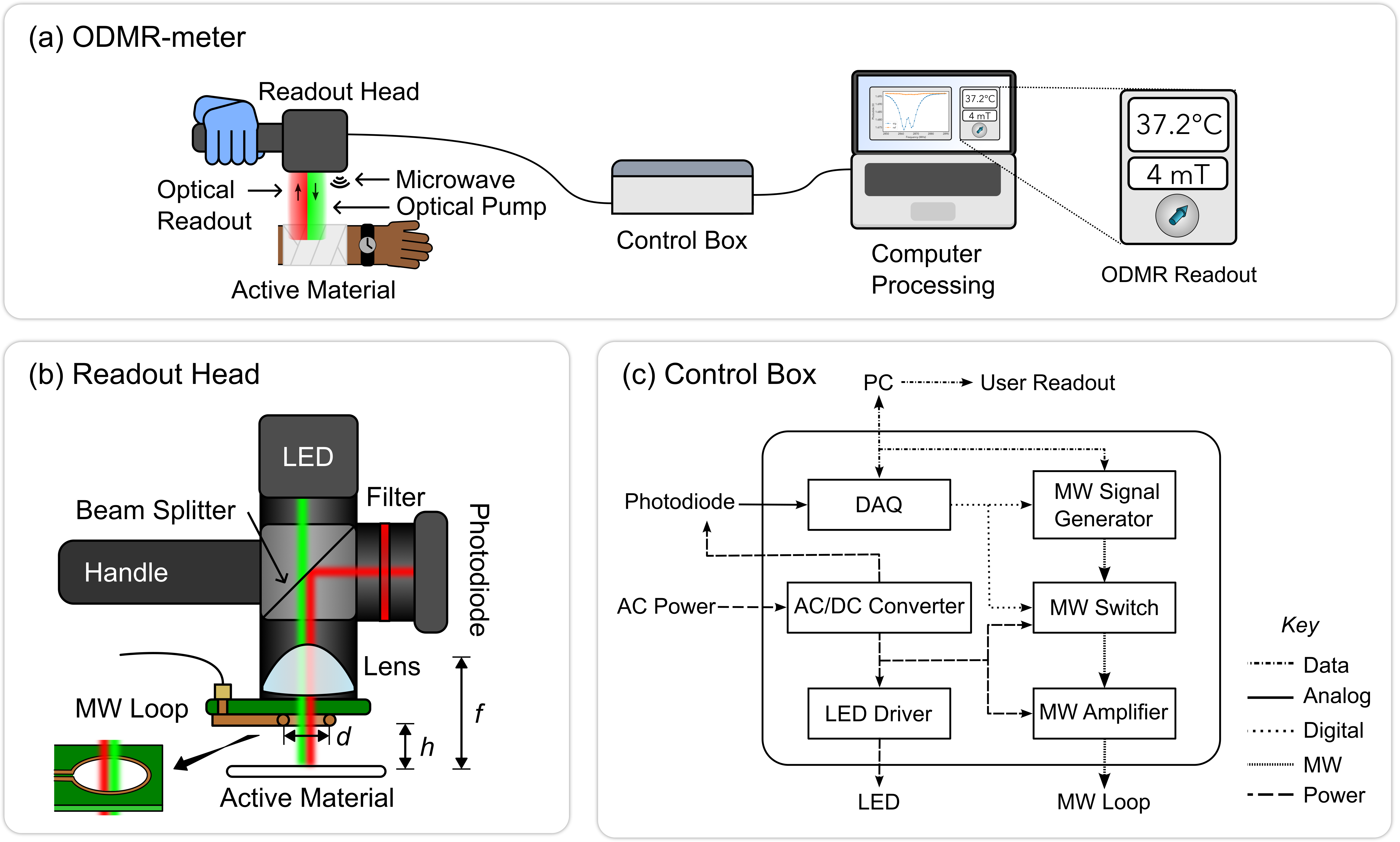}
    \caption{(a) Concept of the ODMR-meter. The central component is a handheld readout head which is aimed at an object to perform ODMR spectroscopy. The readout head is connected to a control box and a computer for data processing and user feedback via a display. (b) Schematic of the readout head which integrates the optical components (including a light-emitting diode, LED) and a microwave (MW) loop antenna. (c) Diagram showing the components included in the control box and their interconnections. DAQ: data acquisition and control unit.}
    \label{fig:device_diagram}
\end{figure*}

The general concept of the ODMR-meter and the various components necessary to its implementation are illustrated in Fig.~\ref{fig:device_diagram}a. The readout head is aimed at an object containing ODMR-active material, and comprises the optical and MW components necessary to perform ODMR spectroscopy of the object at some distance, without physical contact. The readout head is connected to a control box containing the necessary electronics and power supplies. The control box is itself connected to a laptop computer operating the device and displaying the ODMR readout, which can be converted to e.g. temperature or magnetic field.  

Our design for the readout head is detailed in Fig.~\ref{fig:device_diagram}b. In the present implementation, it is designed to measure the photoluminescence (PL) from NV defects in diamond in an epi-illumination geometry, with a green light-emitting diode (LED) for optical excitation, a photodiode to detect the red PL, and a dichroic beam splitter to separate PL from excitation. A condenser lens on the object side serves both to focus the excitation light onto the object and collect the PL. The MWs are supplied via a loop antenna mounted just under the condenser lens, with the light passing through the loop. A key design parameter is the working distance $h$, i.e. the shortest distance between readout head and object. The various considerations that dictate the choice of $h$ are discussed in Sec.~\ref{sec:head}.

As depicted in Fig.~\ref{fig:device_diagram}c, the control box contains a set of MW components (signal generator, switch, amplifier) connected to the MW loop, a data acquisition and control unit (DAQ) which reads the voltage from the photodiode and control the MW signal generator (for MW frequency sweeping) and switch (for amplitude modulation), and power converters to supply power to the LED, photodiode and MW switch. The personal computer (PC) is connected via a universal serial bus (USB) to the DAQ and MW signal generator. A software interface controls the data acquisition and analysis process and displays the outcome.

\subsection{Design of the readout head} \label{sec:head}

The purpose of the ODMR-meter is to measure ODMR from an object at a distance, without contact, by simply holding the readout head above the object. Therefore, it is desirable for the working distance $h$ to be as large as possible, at least several millimeters, ideally centimetres. However, this desire competes with the requirement of a short working distance to optimise device performance. Indeed, a long working distance places limits on the efficiency of the MW driving, optical pumping, and PL collection, which in turn affects the measurement sensitivity and speed -- defined as the maximum rate at which successive readings (e.g. of temperature) can be made. 

We first consider the relationship between working distance and MW driving efficiency. The latter is characterised by the Rabi frequency which is proportional to the strength of the magnetic field associated with the MW, $B_{\rm MW}$. If the object is placed on the symmetry axis of the MW loop, the field it experiences scales with the geometrical parameters as
\begin{align} \label{eq:MW}
B_{\rm MW} &\propto \frac{d}{d^2+4h^2}
\end{align} 
where $d$ is the diameter of the loop. For a given current in the loop and a given working distance $h$, the field is maximised for a loop diameter $d=2h$ and then scales inversely with $h$, $B_{\rm MW}\propto h^{-1}$. Note that a different antenna geometry, for instance a straight wire, would generally give the same scaling with $h$. The MW power to be supplied to the loop then scales as $P_{\rm MW}\propto B_{\rm MW}^2 \propto h^{-2}$. Thus, a working distance of $h=1$~cm requires 100 times more MW power to achieve the same driving efficiency as at $h=1$~mm. In our implementation, we will use $h=4$~mm and $d\approx2h$, which we found (see Sec.~\ref{sec:loop}) to provide good MW driving (i.e. with a Rabi frequency roughly matching the dephasing rate of the NV spins in our test diamond) at the maximum MW power that can be delivered by the chosen hardware. 

Longer working distances can be achieved by scaling the supplied MW power accordingly, however, it also places constraints on the optical design and affects measurement performance. On the one hand, the PL collection efficiency directly affects the measurement sensitivity. Indeed, the sensitivity (to temperature or magnetic field) scales as $\eta\propto I_{\rm PL}^{-1/2}$ where $I_{\rm PL}$ is the collected PL intensity (as detected by the photodiode, assuming shot noise limited measurements)~\cite{Rondin2014}. The PL intensity in turns primarily depends on the numerical aperture (NA) of the collection lens, $I_{\rm PL}\propto{\rm NA}^2$, such that $\eta\propto{\rm NA^{-1}}$. For a given NA, the larger the working distance, the larger the diameter of the lens. Aspheric condenser lenses with NA exceeding 0.7 (about $45^\circ$ acceptance angle) are available with diameters up to 2" or more. For instance, a working distance of $h=4$~mm can be accommodated using a high-NA 1" lens (e.g. Thorlabs ACL25416U, focal length $f=16$~mm) mounted just above the MW loop (accounting for the finite thickness of the lens and MW loop), while a larger 2" lens (e.g. Thorlabs ACL50832U, $f=32$~mm) allows for $h=12$~mm. In our implementation, we chose the $f=16$~mm 1" lens to keep the readout head relatively compact. Note that ${\rm NA}=0.7$ roughly corresponds to an aperture of $d=2h$ at a distance $h$ from the object, such that the MW loop is not restricting light collection in these conditions. If we restrict ourselves to 1" diameter lenses, longer working distances are possible but at the expense of a reduced NA (larger $f$) and thus degraded sensitivity. For instance, $h=8$~mm can be achieved with a 1" lens for a slightly reduced NA of 0.6 (e.g. Thorlabs ACL2520U, $f=20$~mm), and a working distance exceeding a centimetre can be accommodated with an NA below 0.5 (e.g. Thorlabs LA1252, $f=25$~mm).

On the other hand, the choice of lens also affects the focusing of the LED light onto the object. For given LED characteristics (in particular, emitter size), the minimum size of the illumination spot on the object is proportional to the focal length of the lens, $f$. The power density then scales as $I_{\rm LED}\propto f^{-2}$. Thus, for a given LED power a larger $f$ implies a smaller pumping rate of the NV spins (e.g. by a factor of 2.5 going from $f=16$~mm to $f=25$~mm), which can degrade sensitivity in different ways as well as limit the measurement speed~\cite{Dreau2011,Levine2019,Barry2020,Sturner2021}, as will be discussed further in Sec.~\ref{sec:speed}. Moreover, a large illumination spot may reduce the ability to fully collect the PL emission as optical aberrations become more severe. A solution is to use a laser instead of an LED to allow for better control on the focusing, as often employed in realisations of NV magnetometers~\cite{Sturner2021,Wang2022}. Here we opted for an LED as they are widely available and provide higher output power than standard laser-based alternatives. A higher excitation power in turn means a larger total PL intensity from the active material, which is important for a device destined to work in ambient light conditions.

Summarising, increasing the working distance generally degrades the performance of the ODMR-meter (measurement sensitivity and speed), and/or implies more stringent hardware requirements such as higher MW power and optical excitation power. In the next section, we describe a realisation of the ODMR-meter with a working distance of $h=4$~mm, which allows for a simple design with relatively compact hardware. Nevertheless, longer working distances of the order of 1~cm are achievable with standard hardware albeit less compact and portable (e.g. a high-power MW amplifier).

\section{Implementation}

\begin{figure*}[tb!]
    \centering
    \includegraphics[width=18cm]{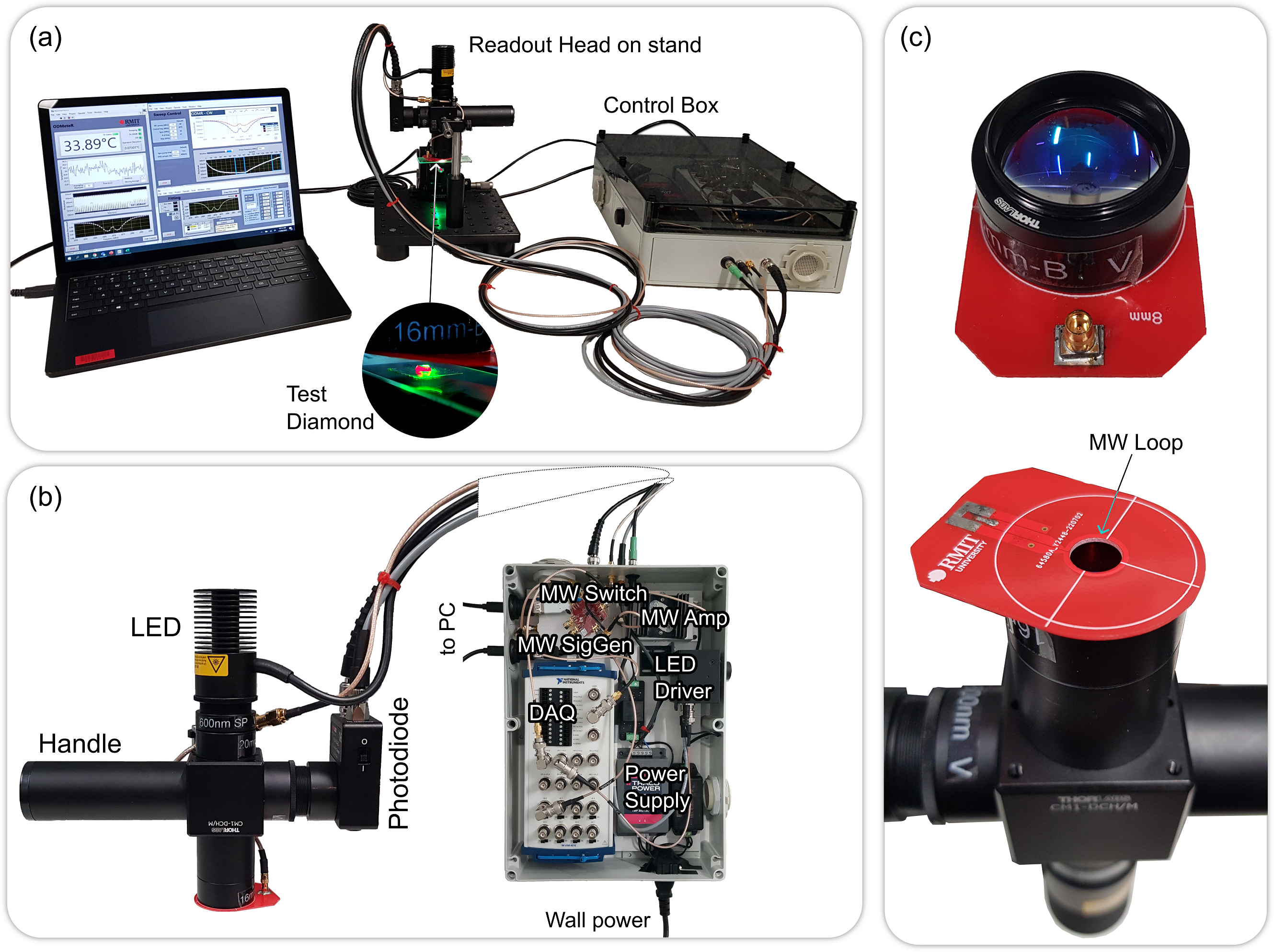}
    \caption{(a) Photograph of the overall apparatus. Here the readout head is held by a stand with a diamond placed below for testing purpose. (b) Photographs of the readout head viewed from the side (left), and of the inside of the control box (right). (c) Photographs of the lens-PCB assembly when separated from the readout head (top), and of the readout head viewed from the object side (bottom).}
    \label{fig:photos}
\end{figure*}

\subsection{Hardware details}

Here we describe the specific details of our prototype. The overall apparatus is photographed in Fig.~\ref{fig:photos}a, with the sub-units shown in Fig.~\ref{fig:photos}b. The readout head is built entirely from Thorlabs components except for the MW loop. A green LED (M530L4) outputs 480 mW power (typical) at the maximum current, centred at 522 nm wavelength (30 nm full width at half maximum intensity). The emitted light is filtered (FESH0600) to remove long wavelengths and collimated by an aspheric condenser lens (ACL2520U-A). The LED-filter-lens assembly (excitation arm) is connected to a cage cube (CM1-DCH) containing a shortpass dichroic beam splitter (DMSP605R). A lens tube (SM1L40) is connected to a port of the cage cube to serve as a handle. The detection arm connected on the output port of the cage cube comprises a longpass filter (FELH0650), an aspheric condenser lens (ACL25416U-B), and an amplified Si photodiode (PDA36A2). On the remaining port of the cage cube, we connect an aspheric condenser lens (ACL25416U-B) and glue a printed circuit board (PCB) directly on the lens-holding tube (SM1L05). Another lens tube (SM1L15) is inserted in between the cage cube and lens to increase physical distance between the handle and the target object. The PCB (shown in Fig.~\ref{fig:photos}c) includes a loop antenna to supply the microwave to the active material, with a hole in the middle to let the light through, and an MMCX connector to connect the MW loop to the control box. The focal point of the lens is located approximately 4~mm below the surface of the PCB, corresponding to the working distance $h$. We tested different loop diameters $d$ and found an optimum at $d\approx9$~mm, see discussion in Sec.~\ref{sec:loop}. All the results shown below have been obtained with this optimum. 

The readout head is connected to the control box via a bundle of four cables (for LED drive, MW drive, photodiode power supply, photodiode output). The largest and heaviest component in the control box is the DAQ (NI USB-6212 BNC). The enclosure (Spelsberg TK PS Series, RS) is 36 x 25 x 11 cm$^3$ in size. In future, the size and weight of the control box could be significantly reduced by using a compact alternative to the DAQ such as an Arduino microcontroller board, which has been employed in previous NV magnetometry demonstrations~\cite{Sanchez2020,Mariani2022}. This could also remove the need for a computer, as the Arduino is capable of processing data and outputting the result to a standalone display. The MWs are generated by a USB-powered signal generator (Windfreak SynthNV Pro) and passed through an IQ modulator used as an on/off switch (Texas Instruments TRF37T05EVM) and an amplifier (Mini-Circuits ZRL-3500+). The control box is connected to AC wall power and includes a set of AC/DC converters (15 V, $\pm12$ V, 3.3 V) to power the LED driver (Thorlabs LEDD1B, housed in the box), photodiode, MW amplifier, and IQ modulator. The total power consumption of the box (including USB inputs) is about 30~W in typical operating conditions.  

\subsection{ODMR acquisition procedure} \label{sec:ODMR}

\begin{figure*}[tb!]
    \centering
    \includegraphics[width=18cm]{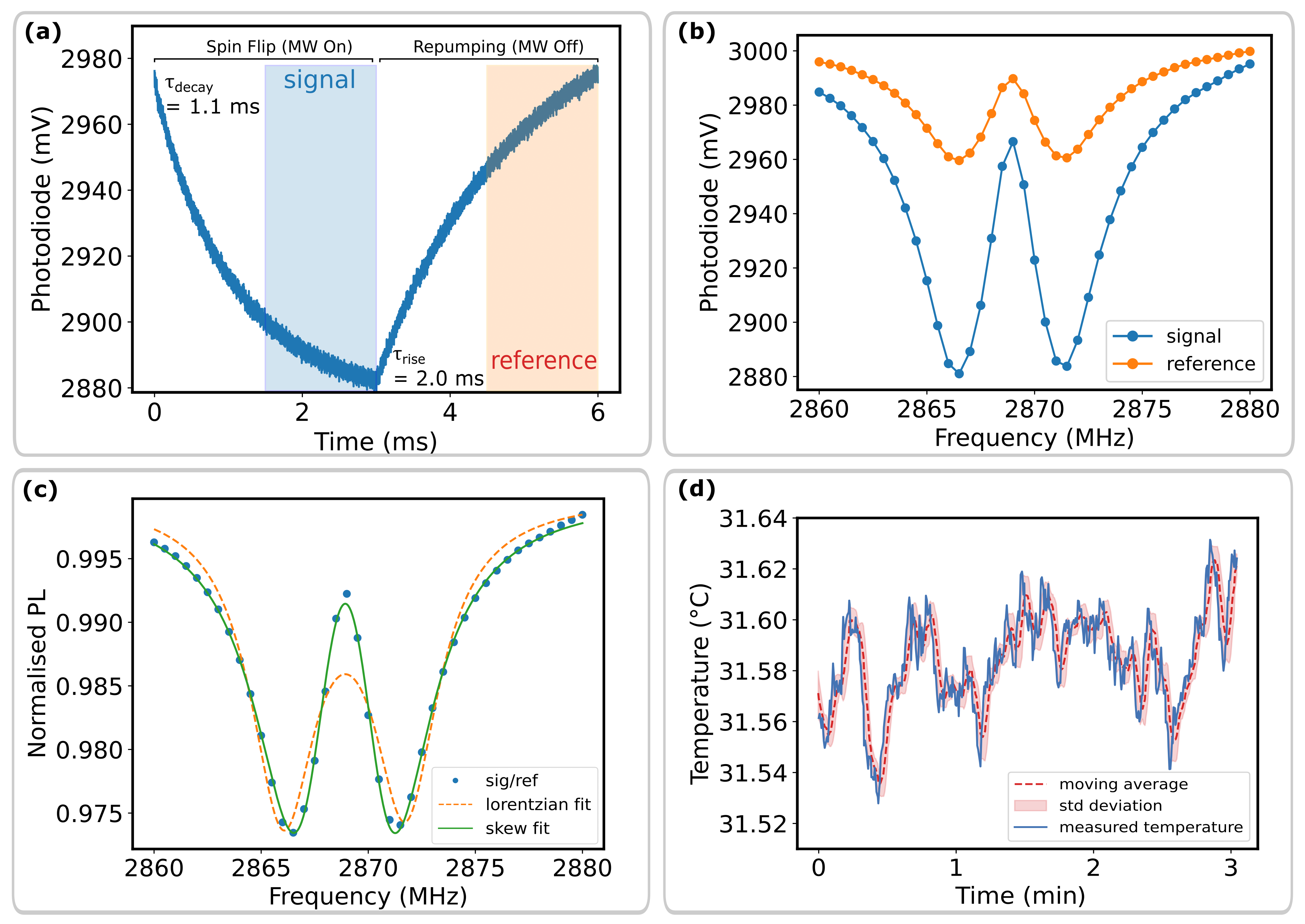}
    \caption{(a) Time trace of the photodiode voltage for a test bulk diamond with the MW frequency tuned to an NV electron spin resonance (2867 MHz). One modulation cycle is shown, with the MW turned on for the first 3 ms (spin-flip phase), and then turned off for the following 3 ms (repumping phase). The shaded areas indicate the regions over which the photodiode signal is averaged to generate the ODMR spectrum. (b) Averaged photodiode signal as a function of MW frequency during a sweep, separating the MW-on (blue data, `signal') and MW-off (orange, `reference') cases. (c) Normalised ODMR spectrum obtained by taking the ratio of the two data sets in (b). The solid lines are fit using Lorentzian (orange line) or skew normal distribution (green) line shapes. (d) Time trace of the temperature estimated from fitting a series of ODMR spectra acquired consecutively. Each data point corresponds to a 250-ms ODMR sweep. The red dashed line is the 10-point moving average; the shaded area corresponds to $\pm1\sigma$ where $\sigma$ is the standard deviation calculated from the last 10 measurements.}
    \label{fig:data}
\end{figure*}

A standard laptop computer and a LabVIEW interface are used to control the DAQ and MW generator to continuously acquire ODMR spectra, as well as to process the data and display the output in real time. To reduce the impact of low-frequency fluctuations (of LED power, PL intensity, MW driving efficiency), which is especially important for handheld operation, a lock-in detection technique is employed. Here we use amplitude modulation, using the MW switch to turn the MW on and off with the LED kept on at all times. The modulation frequency is limited by the repumping rate of the NV spin, in turn limited by the intensity of the green light~\cite{El2017,Deng2020}. We use a modulation frequency in the range $f_{\rm mod}=100-500$~Hz, e.g. at a typical value of $f_{\rm mod}=167$~Hz the MW is turned on for 3~ms (spin-flip phase) and turned off for 3~ms (repumping phase). To acquire an ODMR spectrum, the MW frequency is then swept across the expected NV electron spin resonances near 2870 MHz, with one modulation cycle per frequency step. To optimize measurement speed and sensitivity, it is possible to use a small set of carefully chosen frequency steps~\cite{Kucsko2013,Fujiwara2020,Fujiwara2020b}. However, for the tests performed here we will use a linear ramp over a relatively wide range, e.g. from 2860 to 2880 MHz with 41 frequency steps (i.e. 0.5 MHz increments), which gives a full well-resolved ODMR spectrum allowing reliable data fitting. In this case, a single ODMR sweep thus takes about 250 ms.

Meanwhile, the voltage from the photodiode is sampled by the DAQ at the maximum available rate of 400 kHz, which gives 1200 samples per 3 ms period. An example time trace of a single MW on/off cycle (6 ms, $f_{\rm mod}=167$~Hz) is shown in Fig.~\ref{fig:data}a, with the MW frequency tuned on resonance. For the tests presented in this section and throughout the paper, we used a type-Ib diamond (Element Six, $3\times3\times0.3$~mm$^3$ in size) that was electron irradiated and annealed to produce a high density of NV defects~\cite{Acosta2009}, and the readout head was held in a fixed position above the diamond using a stand as shown in Fig.~\ref{fig:photos}a. The trace in Fig.~\ref{fig:data}a illustrates the effect of the MW modulation: when the MW is turned on, the PL intensity decays exponentially with a characteristic decay time of $\approx1.1$~ms (spin-flip time, dependent on the MW field strength and optical excitation rate); when the MW is turned off, the PL follows an exponential recovery with a characteristic time of $\approx2.0$~ms (repumping time, dependent on the optical excitation rate). The raw DAQ data is processed by first averaging the samples from the final part of each decay, as indicated by the shaded areas in Fig.~\ref{fig:data}a. This averaged signal is plotted against MW frequency in Fig.~\ref{fig:data}b for both the MW-on case (called `signal') and the MW-off case (`reference'). By dividing the signal by the reference, we obtain the normalised (demodulated) spectrum, shown in Fig.~\ref{fig:data}c. Note that the reference spectrum still exhibits resonance features, which is due to the modulation being too fast to allow complete spin repumping, causing a reduction in contrast in the normalised ODMR spectrum (nearly 3\% here). A slower modulation can increase the contrast up to 6\% but at the cost of a reduced measurement speed -- it takes longer to complete one sweep. Reducing $f_{\rm mod}$ will also eventually compromise the sensitivity since it reduces the rate of collected PL for a given averaging window (shaded areas in Fig.~\ref{fig:data}a). Here, $f_{\rm mod}$ was chosen to maximise the signal-to-noise ratio in the normalised ODMR spectrum, while optimising the width of the averaging window for each value of $f_{\rm mod}$. This optimisation resulted in a minimum standard deviation in the resulting temperature trace, discussed below. 

\subsection{Temperature estimation} \label{sec:thermometry}

In normal operation, ODMR spectra are recorded continuously, i.e. a new spectrum like that in Fig.~\ref{fig:data}c is produced every 250 ms in our example. In general each spectrum can be fitted by LabVIEW in a time much shorter than this single-sweep time (about 10~ms to fit a full spectrum, which is done in parallel to the acquisition of the next sweep), thus allowing real-time feedback to the user. Here we fit the spectrum with two lines corresponding to the two effective electron spin resonances expected for an ensemble of NV defects in zero magnetic field~\cite{Mittiga2018}. These two resonance frequencies can be expressed as $f_{\pm}=D\pm E$ where $D$ and $E$ are the zero-field splitting parameters. 

The $D$ parameter depends on temperature, with a linear coefficient of $\frac{{\rm d}D}{{\rm d}T}\approx-75$~kHz/K near room temperature~\cite{Acosta2010}. For better accuracy (see further discussion in Sec.~\ref{sec:accuracy}), we use the second-order polynomial
\begin{align} \label{eq:T}
D(T) &= a_0 + a_1 T + a_2 T^2
\end{align}
with coefficients $a_0=2.875357$~GHz, $a_1=4.3998\times10^{-5}$~GHz/K and $a_2=-2.0519\times10^{-7}$~GHz/K$^2$. This polynomial is based on a fit to the third-order polynomial given in Ref.~\cite{Toyli2012} over the range 0-100$^\circ$C relevant to our application. The temperature can thus be estimated by solving for the positive root of $f(T) = D(T) - D_{\rm meas}$ where $D_{\rm meas}$ is the value estimated from the ODMR fit. 

To obtain visually good fitting, we use skew normal distributions with skew parameters of identical amplitude but opposite signs for the two resonances, resulting in the green line in Fig.~\ref{fig:data}c. Symmetric lineshapes fail to capture the narrow central peak, see e.g. the Lorentzian fit in Fig.~\ref{fig:data}c (orange line). The asymmetric lineshape is understood to be due to the interplay between local random electric fields and magnetic noise~\cite{Mittiga2018}. While it is possible to fit the resulting spectrum using a physical model~\cite{Mittiga2018}, this is a computationally intensive task (the fitting time would exceed the single-sweep time) and therefore we prefer to use a simple skew normal distribution. The implications of this choice on the measurement accuracy will be discussed in Sec.~\ref{sec:accuracy}.

By analysing each single-sweep spectrum using a skew normal fit and deducing the temperature via Eq.~\ref{eq:T}, we obtain a time trace of the temperature sampled every $\approx250$~ms (the single-sweep time), shown in Fig.~\ref{fig:data}d. The typical standard deviation calculated from series of 10 samples is 20~mK, which corresponds to a sensitivity of $\eta_T\approx10~{\rm mK}/\sqrt{\rm Hz}$. This directly measured value is close to the sensitivity limit set by the noise in the photodiode signal (visible in Fig.~\ref{fig:data}a) in the case of an optimised single MW frequency measurement (chosen in the steepest part of the ODMR spectrum), of about $\approx7~{\rm mK}/\sqrt{\rm Hz}$. Note, if the noise in the photodiode signal was purely due to photon shot noise, given an estimated detected PL power of about 1.8 mW, the sensitivity would be improved to $\approx1~{\rm mK}/\sqrt{\rm Hz}$. The limits and tradeoffs to sensitivity will be discussed in Sec.~\ref{sec:sensitivity}.

\subsection{MW loop optimisation} \label{sec:loop}

One important design parameter is the diameter of the MW loop, $d$. For a given working distance $h$, the MW field strength at the object is maximised for a diameter $d=2h$, according to Eq.~\ref{eq:MW}. However, as the loop itself partially blocks light focusing and collection by the lens, it is useful to determine the optimal diameter experimentally. To this end, we compared PCBs with different loop diameters from $d=5.0$~mm to $d=13.4$~mm. The central hole in the PCB is as large as permitted by manufacturing tolerance; for instance a hole diameter of 8 mm corresponds in fact to a mean loop diameter of $d=9.4$~mm (accounting for the finite width of the strip line forming the loop). For each value of $d$, we placed the test diamond at the working distance of $h=4$~mm (set by the lens) and recorded an ODMR spectrum under fixed conditions (in particular, using the maximum LED and MW powers available). We then extract the PL intensity ($I_{\rm PL}$, measured off resonance) and, by fitting the spectrum, the contrast ${\cal C}$ and linewidth $\Delta\omega$ (full width at half maximum) of the resonances. 

\begin{figure}[tb!]
    \centering
    \includegraphics[width=8cm]{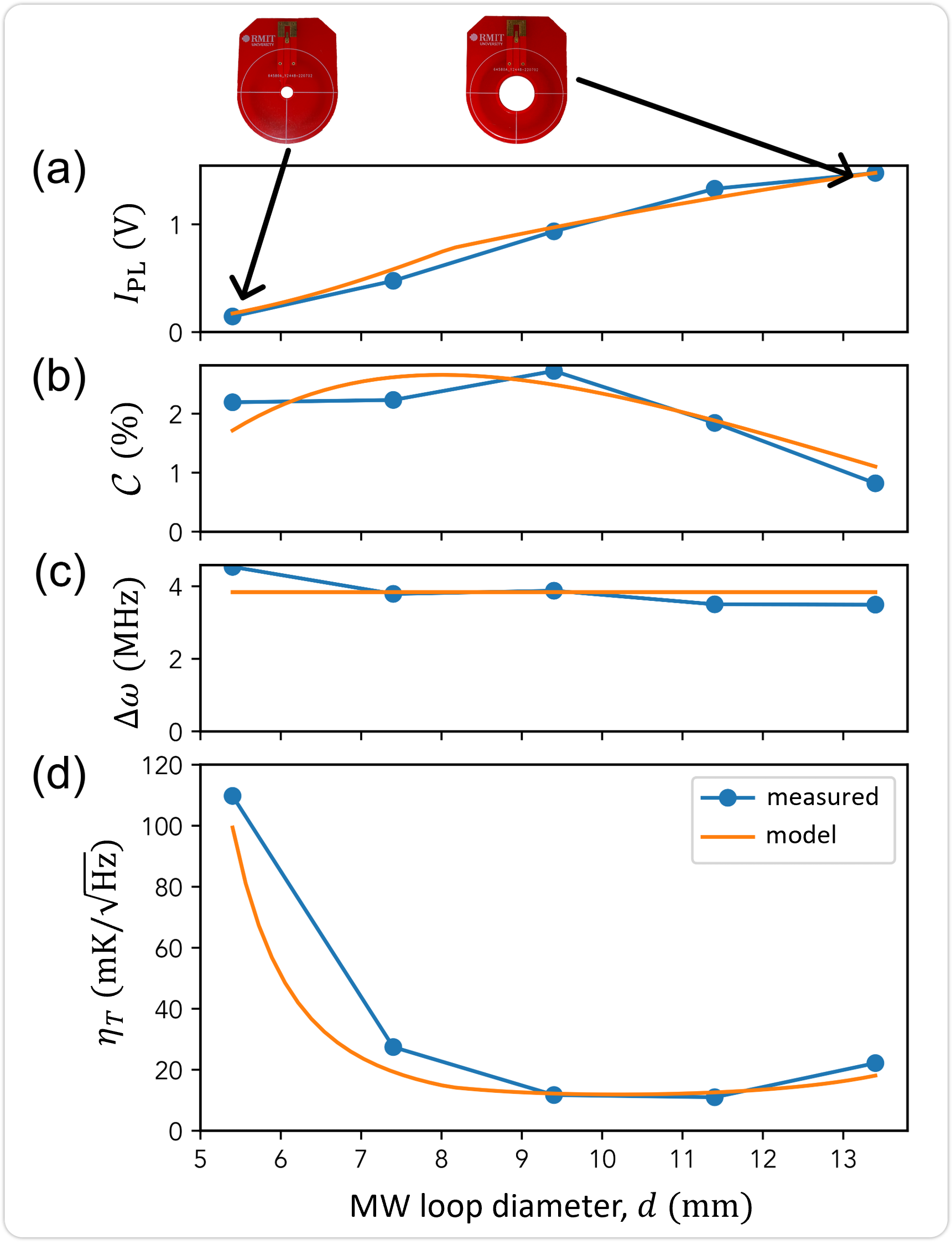}
    \caption{(a) PL intensity measured by the photodiode as a function of the MW loop diameter. The photographs at the top show the PCB for the smallest and largest diameters. Note, the gain setting of the photodiode was different to that in Fig.~\ref{fig:data}, resulting in a lower voltage. (b,c) Contrast (b) and linewidth (c) of the resonances estimated by fitting the ODMR spectrum. (d) Sensitivity $\eta_T$ estimated from Eq.~\ref{eq:eta} using the data in (a-c), with the pre-factor $\beta$ obtained considering the experimentally observed noise for these measurements. In (a-d), the orange solid lines are the model with free pre-factors to fit the data, see text for details.}
    \label{fig:design_tradeoffs}
\end{figure}

The three quantities $I_{\rm PL}$, ${\cal C}$, and $\Delta\omega$, are plotted against loop diameter in Fig.~\ref{fig:design_tradeoffs}a-c, respectively. The orange solid lines are the result of simple theoretical considerations. For the PL intensity, we consider clipping by the aperture in the PCB of both the excitation light from the LED and the collected PL (see modelling details in Appendix~\ref{App:model}). The predicted nearly linear trend agrees well with experiment (Fig.~\ref{fig:design_tradeoffs}a, an arbitrary pre-factor is adjusted to best fit the data). For the contrast (Fig.~\ref{fig:design_tradeoffs}b), the model is based on Eq.~\ref{eq:MW}, with a pre-factor left as a fit parameter, assuming a linear relationship between ${\cal C}$ and $B_{\rm MW}$ approximately valid in this regime of low excitation powers~\cite{Dreau2011}. The data agrees with the expectation of a maximum contrast near the $d=2h$ condition. For the linewidth (Fig.~\ref{fig:design_tradeoffs}c), we expect a negligible variation since here $\Delta\omega$ is close to the limit set by the spin dephasing rate, $1/T_2^*$~\cite{Dreau2011}; the data indeed shows a weak dependence with the loop diameter. 

In the present case where the noise given by the photodiode is a constant independent of signal amplitude (i.e. $I_{\rm PL}$), the temperature sensitivity scales inversely with the maximum slope in the ODMR spectrum and can thus be expressed as
\begin{align} \label{eq:eta}
\eta_T=\beta\frac{\Delta\omega}{{\cal C}I_{\rm PL}},    
\end{align}
where the pre-factor $\beta$ characterises the (constant) noise amplitude. The sensitivity thus estimated is plotted against $d$ in Fig.~\ref{fig:design_tradeoffs}d, where the data points are calculated using the values measured in Fig.~\ref{fig:design_tradeoffs}a-c, whereas the solid line uses the models described above for each quantity in Eq.~\ref{eq:eta}. The predicted minimum at $d\approx10$~mm is in good agreement with the experiment, which finds a best sensitivity for the PCB with a MW loop diameter $d=9.4$~mm and an 8-mm hole. 

In summary, the optimal MW loop diameter is found to be close to that given by maximising the MW field strength according to Eq.~\ref{eq:MW}, with a small correction (a larger diameter by about 20\%) to allow for more light to go through the hole. This simple design rule can serve to adjust the PCB design for different working distances. 

\section{Demonstration of handheld operation} \label{sec:handheld}

\begin{figure*}[tb]
    \centering
    \includegraphics[width=14cm]{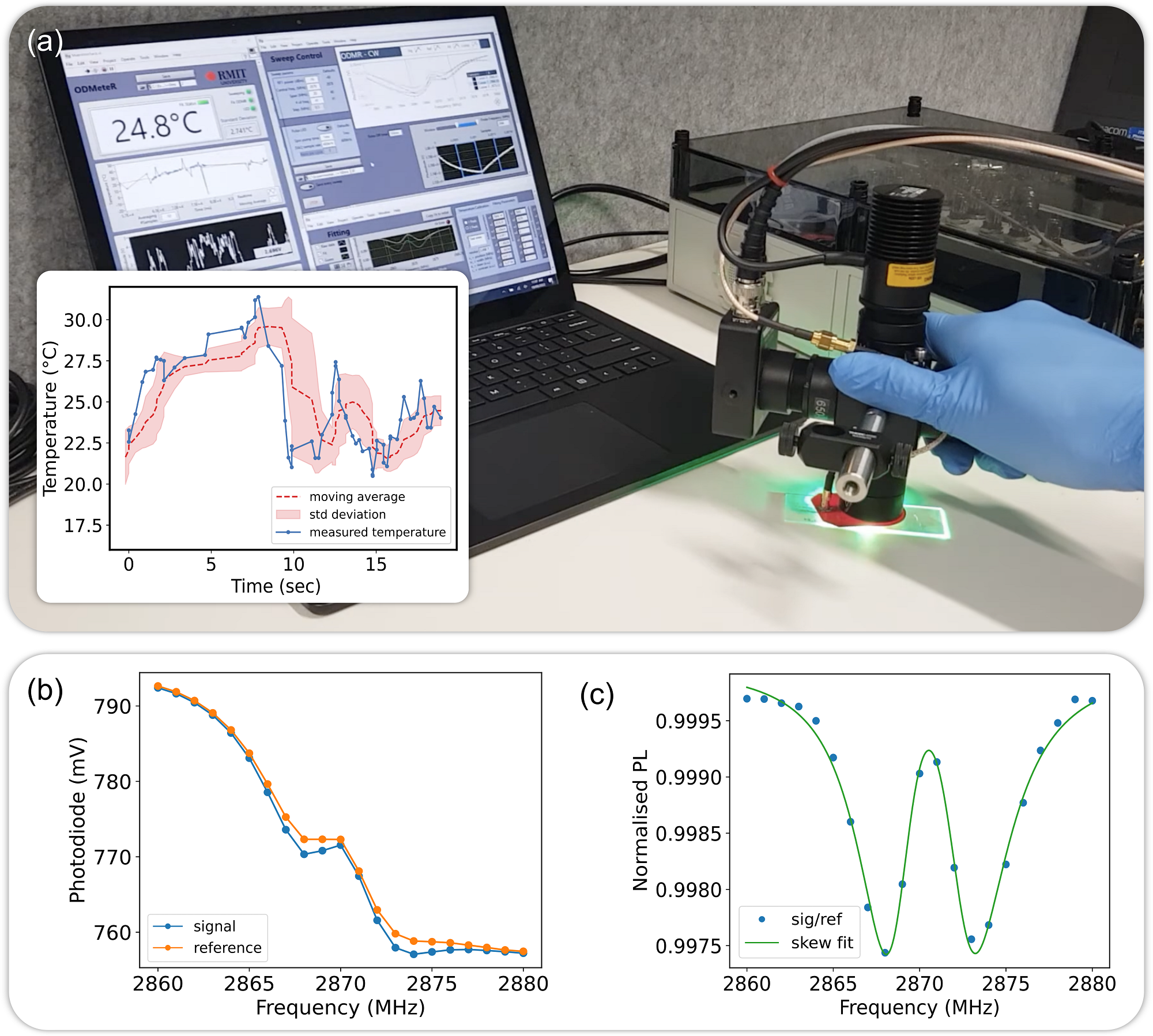}
    \caption{(a) Photograph of the setup to test handheld operation. Inset: example time trace of a 20-second burst of continuous temperature readings during handheld operation. (b) Single ODMR sweep showing the raw signal and reference traces, obtained with the readout head held by hand above a test diamond. (c) Normalised spectrum obtained by taking the ratio of the two traces in (b). The solid line is the skew normal fit.}
    \label{fig:video}
    % Params
    % 1ms pump time (2ms total)
    % -20 dBm RF
    % 20 MHz span
    % 41 points
    % video: TrimmedSensing.mp4
\end{figure*}

The results presented so far were obtained with the readout head placed on a stand with a fixed position relative to the target object, as shown in Fig.~\ref{fig:photos}a. However, some applications will require the readout head to be held by hand, for example to get a rapid temperature reading of a smart dressing applied to an immobile object. Alternatively, the ODMR-meter could be used in clinical settings where the target is not immobilised, for instance a smart dressing applied to a patient or an animal model. In the latter case, assuming the animal is placed under anaesthesia, the readout head could be mounted on a stand, but the breathing movement of the animal would still result in a time-dependent standoff distance between readout head and object. 

To test the ability of the ODMR-meter to perform measurements under such dynamic conditions, we placed our test diamond on a table and held the readout head by hand, as shown in Fig.~\ref{fig:video}a. The ODMR sweep is running continuously, but initially shows no contrast as the diamond is not near the focal point of the lens. As the sensor head is brought near the diamond, the photodiode signal increases and the resonances become visible in the ODMR spectrum. An estimate of the temperature is displayed whenever the fit converges with a residual below a user-defined threshold. A video of this process is available as supplementary online material, illustrating the intermittent nature of the measurement with bursts of valid readings for several seconds in a row separated by periods with no valid reading while the operator is repositioning the head. An example time trace of the temperature during a continuous 20-second burst is shown in the inset of Fig.~\ref{fig:video}a. For these measurements, we increased the modulation frequency to $f_{\rm mod}=500$~Hz (which reduces the contrast to $<1\%$) and reduced the number of frequency steps in the ODMR sweep to 21, so that one sweep takes 42 ms. We found this sweep time to be sufficiently fast to make temperature reading relatively easy despite the unavoidable movement of the holding hand. An example ODMR spectrum thus obtained is shown in Fig.~\ref{fig:video}b (raw traces) and Fig.~\ref{fig:video}c (normalised spectrum). The raw traces show a large common-mode modulation due to the physical movement, which is largely removed in the normalised spectrum, although we note that the inferred temperature exhibits increased fluctuations (see Fig.~\ref{fig:video}a inset) compared with the static case. This experiment validates the feasibility of operating the ODMR-meter by hand to measure the temperature of an object.

\section{Performance limitations}

Having demonstrated successful operation of the ODMR-meter in both static and dynamic handling conditions, we now discuss the current limitations to its performance, and possible improvements.

\subsection{Measurement speed} \label{sec:speed}

Measurement speed here refers to the maximum rate at which successive temperature readings can be made, and is characterised by the minimum time $t_{\rm read}$ required to get one reading of the temperature. This is an important quantity in the context of a handheld device because $t_{\rm read}$ needs to be short enough to allow one correct reading before moving out of focus. In Sec.~\ref{sec:handheld}, we saw that a reading time of $t_{\rm read}=42$~ms was sufficient to get one reading when holding the readout head by hand above a fixed object. However, operation may be facilitated by a shorter measurement time, especially in a less controlled environment (leading to a less steady hand) or if the object itself is moving. While we used 21 frequency points in our demonstration (2~MHz constant spacing between points), it is generally accepted that 4 carefully chosen frequency points are sufficient to enable an accurate estimate of the temperature~\cite{Kucsko2013,Fujiwara2020,Fujiwara2020b}. Using the same modulation frequency of $f_{\rm mod}=500$~Hz, i.e. 2~ms per frequency step, this 4-point ODMR measurement would correspond to a reading time of $t_{\rm read}=8$~ms. 

The modulation period, i.e. the time per frequency step, is limited by the repumping rate, which has a characteristic time of $\approx2$~ms in our current implementation according to Fig.~\ref{fig:data}a. This repumping time depends on the optical excitation intensity and the NV absorption cross-section. We measured the excitation power incident on the diamond to be about 160 mW, spread over an area of roughly 3~mm$^2$, i.e. a power density of $\sim5$~W/cm$^2$, consistent with the observed repumping time~\cite{Sturner2019}. Increasing the repumping rate requires to increase the optical power, or more feasibly to reduce the excitation area. The latter can be achieved by employing a laser instead of an LED, with characteristic repumping times as short as a few microseconds achievable with tightly focused laser illumination~\cite{Dreau2011}. Consequently, a reading time of $t_{\rm read}\sim10-100~\mu$s is in principle achievable. Note that there may be other hardware limitations, for instance in our case the MW signal generator used has a settling time in frequency sweeping that limits the modulation to $f_{\rm mod}\approx4$~kHz hence a minimum reading time of $t_{\rm read}\approx1$~ms. Such a modulation frequency is achievable using laser excitation even in a compact implementation~\cite{Sturner2021}.    

However, the temperature resolution of a single reading, defined as the minimum detectable temperature change, $\delta T_{\rm min}$, will increase with decreasing reading time according to $\delta T_{\rm min}=\eta_T/\sqrt{t_{\rm read}}$ where $\eta_T$ is the sensitivity introduced before. Thus, depending on the required resolution for a given application, it may be necessary to use a long reading time or average multiple consecutive readings, such that maintaining mechanical stability over the course of a measurement will still be critical. We will next look at how the sensitivity $\eta_T$ itself can be improved.

\subsection{Measurement sensitivity} \label{sec:sensitivity}

If the ODMR lineshape remains unchanged apart from the frequency shift induced by a temperature change, then the sensitivity $\eta_T$ depends only on the maximum slope in the ODMR spectrum and the noise in that spectrum. An expression for $\eta_T$ is given by Eq.~\ref{eq:eta}, where the pre-factor $\beta$ depends on the origin of the noise (for photon shot noise, $\beta\propto\sqrt{I_{\rm PL}}$). We saw in Sec.~\ref{sec:thermometry} that the sensitivity in our case was an order of magnitude above the photon shot noise limit, due to electronic noise from the photodiode. With a faster sampling rate ($\sim10$~MHz, using a faster DAQ or a proper lock-in amplifier) to render this noise comparatively negligible, it should be possible to obtain a photon shot noise limited sensitivity, which would then scale as~\cite{Dreau2011}
\begin{align} \label{eq:eta2}
\eta_T=\beta'\frac{\Delta\omega}{{\cal C}\sqrt{I_{\rm PL}}}   
\end{align} 
where $\beta'$ encapsulates the physical constants.

In our tests, we used a $3\times3\times0.3$~mm$^3$ type-Ib single crystal diamond, and estimated a sensitivity of $\eta_T\approx1~{\rm mK}/\sqrt{\rm Hz}$ in the photon shot noise limit. For the type of application envisioned for the ODMR-meter (smart dressing for temperature monitoring), rather than a bulk crystal, diamond particles with diameters between $\sim10$~nm and $\sim100~\mu$m are likely to be employed and embedded in a flexible matrix for ease of application to the target object~\cite{Andrich2018,Khalid2020}. Particles made out of type-Ib diamond are commercially available and we can expect to obtain a similar sensitivity to our test bulk diamond if the particles are densely packed within the host matrix. 

It is useful to examine Eq.~\ref{eq:eta2} to identify potential pathways to improve the sensitivity below $\sim1~{\rm mK}/\sqrt{\rm Hz}$. The variable quantities in Eq.~\ref{eq:eta2}, namely the measured PL intensity $I_{\rm PL}$, the ODMR contrast ${\cal C}$, and the ODMR linewidth $\Delta\omega$, depend on the properties of the diamond material used (e.g. density of NVs, spin dephasing rate $1/T_2^*$) and the measurement conditions (optical excitation power density, MW driving strength). All of these parameters should be optimised together following the same prescriptions as for NV magnetometry~\cite{Dreau2011,Rondin2014,Barry2020}, given the constraints set by the application. In particular, for continuous-wave ODMR sensing as performed here, there is an optimal set of optical excitation power density and MW driving strength that minimises the sensitivity for a given dephasing rate~\cite{Dreau2011,Levine2019}. In general, the optimal excitation power density is larger than that employed here ($\sim5$~W/cm$^2$), which further motivates the switch to laser excitation instead of LED and could provide a several-fold improvement in sensitivity. Moreover, diamond grown by chemical vapor deposition with a precisely tunable density of NVs, rather than type-Ib diamond, should lead to further sensitivity improvements~\cite{Barry2020}, but mass production of  optimised diamond particles suitable for incorporation into smart dressings is still an outstanding challenge~\cite{Trusheim2014}. 

The measured PL intensity $I_{\rm PL}$ also depends on the collection efficiency of the system, and on the total excitation power. In our device, we collect the PL via a high-NA condenser lens, and given the non-contact nature of the ODMR-meter design it is unlikely that the collection efficiency can be increased much further. Likewise, the excitation power was about 160 mW supplied by an LED and cannot be easily increased much further. On the contrary, using laser excitation via a laser diode module integrated in the head~\cite{Wang2022} or via a fiber input~\cite{Webb2019,Zheng2020,Sturner2021} may restrict the excitation power to less than that employed here, reducing the benefits of laser excitation over LED.

Finally, for handheld operation, the effect of physical movement of the readout head relative to the object on the measurement sensitivity remains to be characterised. It can be expected that the lock-in detection scheme (with $f_{\rm mod}=500$~Hz as used in Fig.~\ref{fig:video}) will protect against motion at lower frequencies, but further work is needed to fully characterise the effect of movement or other sources of noise in real-world scenarios.    

\subsection{Measurement accuracy} \label{sec:accuracy}

The accuracy of a temperature reading performed by the ODMR-meter depends mainly on two factors that can be seen in Eq.~\ref{eq:T}: (i) the a priori knowledge of the $D(T)$ relationship, via e.g. the polynomial coefficients $\{a_i\}$; and (ii) the accuracy with which the $D$ parameter is determined experimentally. 

The temperature dependence of the NV defect's zero-field splitting parameter $D$ was initially investigated near room temperature by Acosta et al~\cite{Acosta2010}, with a linear coefficient determined to be $\alpha_T=\frac{{\rm d}D}{{\rm d}T}=-75.0(6)$~kHz/K, but the four diamond samples employed in their study exhibited different temperature responses from $-71(1)$~kHz/K to $-78(1)$~kHz/K. Subsequent studies performed over wider temperature ranges fitted the $D(T)$ data with a polynomial~\cite{Chen2011,Toyli2012,Ouyang2021}. 
According to the polynomials given in these three studies (Refs.~\cite{Chen2011,Toyli2012,Ouyang2021}), a reading of $D(T)=2870.00$~MHz translates to a temperature of $T=304.6$~K, $T=301.1$~K, and $T=299.3$~K, respectively. That is, the deduced temperature differs by over 5~K depending on which reference is employed for the conversion. This variability may be due to errors in calibrating the temperature or to actually different responses between diamonds, possibly due to different levels of lattice strain. Therefore, it will be critical to carefully calibrate the $D(T)$ dependence on the diamond material used in the final application, for instance type-Ib diamond microparticles embedded in a smart dressing. With proper calibration, an absolute accuracy below 1~K seems feasible. In Sec.~\ref{sec:thermometry}, we chose to use the third-order polynomial given in Ref.~\cite{Toyli2012} as it gives the most sensible values near room temperature with our test diamond. For simplicity, we fitted this third-order polynomial with a second-order polynomial over the range 0-100$^\circ$C, with resulting coefficients given in Sec.~\ref{sec:thermometry}. With the simplified polynomial, our ODMR-meter reads a minimum temperature of $\approx22^\circ$C under low heating conditions (see next section for details) in agreement with the independently measured room temperature.  

On the other hand, the experimental determination of $D$ from the ODMR spectrum is prone to systematic errors mainly due to asymmetries in the overall lineshape, as studied in detail in Ref.~\cite{Fujiwara2020}. One important source of asymmetry in our implementation is the finite repumping time (set by the modulation frequency $f_{\rm mod}$) which causes a slight slant in the ODMR spectrum due to incomplete spin repumping during the sweep, although the resulting bias can be averaged out by alternating the MW sweep direction. Less trivially, however, any frequency-dependent variation in MW driving efficiency (due to actual change in MW power, or to variations in MW-spin coupling related e.g. to strain-induced spin mixing) can give rise to asymmetries in the ODMR spectrum. Consequently, when fitting the spectrum with two resonances at frequencies $f_\pm=D\pm E$, the value of $D$ returned by the fit generally depends on the lineshape assumed for these resonances, unless they are perfectly symmetric with respect to $D$. For instance, the two fitting methods in Fig.~\ref{fig:data}c give $D$ values that lead to temperatures differing by up to several K in some cases. Importantly, a better visual fit is not a guarantee of accuracy, since the underlying model may be less physically correct. Fitting the data with a complete physical model such as that proposed in Ref.~\cite{Mittiga2018} will likely be necessary to obtain good accuracy while being robust against lineshape variations during operation. This will require careful calibration to determine the parameters of this physical model for the diamond material used in the final application. With a properly parametrised fitting method, one can expect an absolute accuracy below 1~K.  

\subsection{Induced heating}

\begin{figure}[tb]
    \centering
    \includegraphics[width=8cm]{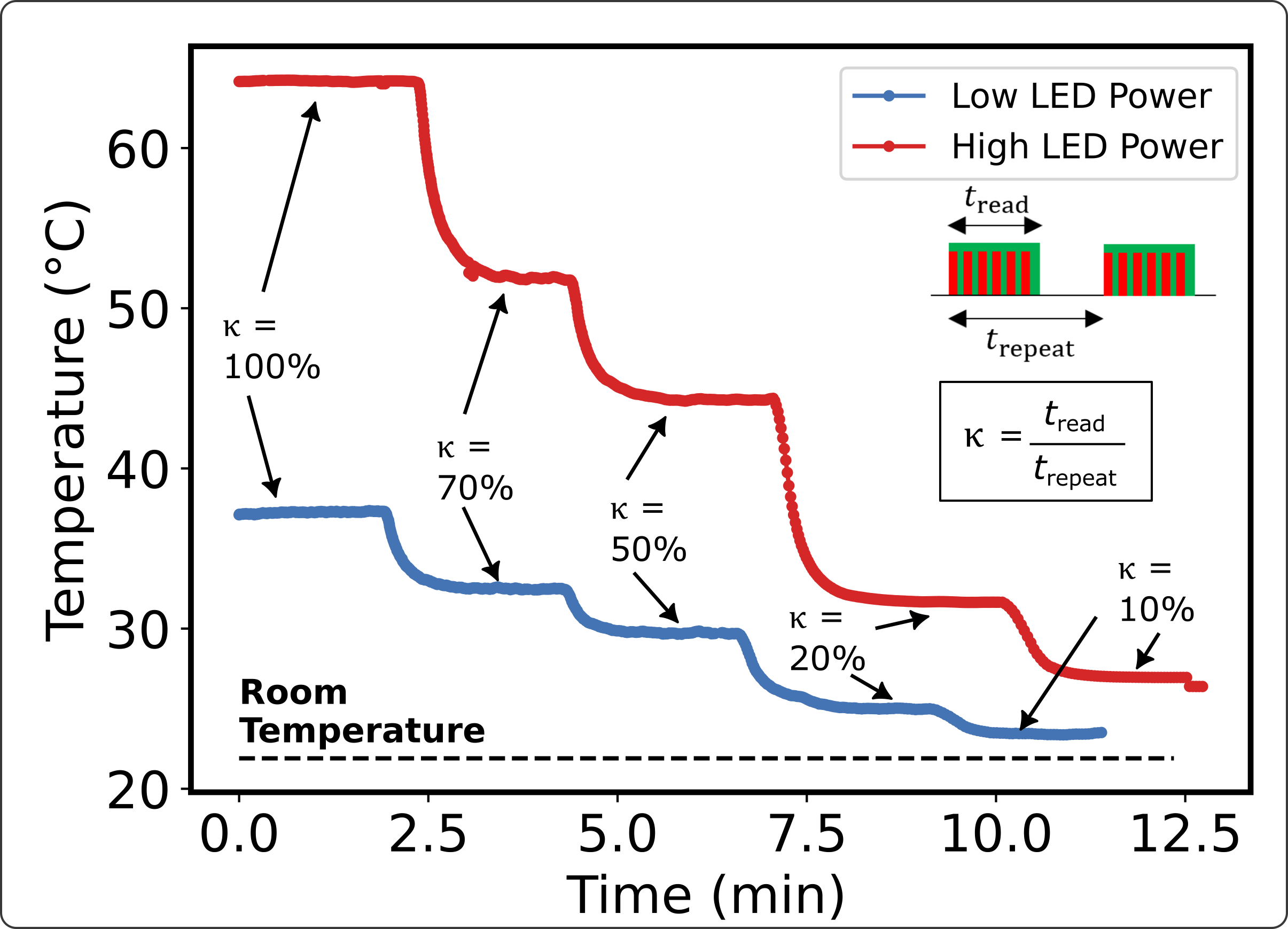}
    \caption{Time traces of the temperature recorded while reducing the duty cycle $\kappa$ in a step-wise fashion, with the LED peak power set to its highest value (160 mW, red trace) or 5 times lower (32 mW, blue trace). The definition of $\kappa$ and the corresponding measurement sequence (green for LED, red for MW) are shown as insets; ODMR sweeps (duration $t_{\rm read}$) are repeated every $t_{\rm repeat}$ with the LED and MW turned off in between. The horizontal dashed line indicates the room temperature.}
    \label{fig:heating}
\end{figure}

A limitation of the ODMR-meter for temperature monitoring applications is that the measurement itself may increase the object's temperature due to the optical as well as MW excitation being partly absorbed by the object. In our tests, we found that optically-induced heating generally dominates, although MW-induced heating also has a measurable effect. For instance, the trace shown in Fig.~\ref{fig:data}d indicates a mean diamond temperature in excess of $31^\circ$C, obtained with the LED set to its maximum output power (160 mW measured at the object), which decreases to $25^\circ$C by reducing the LED power 5 fold while keeping the MW power to its maximum value (200 mW sent to the MW loop). Reducing the MW power by two orders of magnitude further decreases the temperature by less than $1^\circ$C.

Obviously, the object's temperature depends on the heat dissipation pathways available. As an example, by mounting our test diamond more loosely on its support, the diamond temperature reached nearly $70^\circ$C under continuous measurement at the maximum LED and MW powers. To reduce the induced heating without reducing the optical power density and MW driving strength during the measurement, we implemented an intermittent reading scheme whereby LED and MW are turned off for some time between consecutive readings. The average power (optical + MW) incident on the object is thus reduced proportionally to the duty cycle $\kappa=t_{\rm read}/t_{\rm repeat}$ where $t_{\rm repeat}$ is the repetition time (see inset in Fig.~\ref{fig:heating}). The temperature traces in Fig.~\ref{fig:heating} show how reducing the duty cycle dramatically reduces the diamond temperature, from $\approx70^\circ$C at $\kappa=100\%$ to $\approx27^\circ$C at $\kappa=10\%$ with the LED peak power set to its maximum value (160 mW, red trace in Fig.~\ref{fig:heating}). Meanwhile, the sensitivity is degraded by a factor $1/\sqrt{\kappa}\approx3$ for $\kappa=10\%$, and is about $\eta_T\approx30~{\rm mK}/\sqrt{\rm Hz}$ at this duty cycle as estimated from the time trace in Fig.~\ref{fig:heating}. Reducing the LED peak power to 32 mW brings the temperature further down to $\approx24^\circ$C at $\kappa=10\%$. just $\approx2^\circ$C above the room temperature, accompanied by a further reduction in sensitivity. 

In practice, the heating induced by the measurement will need to be characterised in the context of the actual use case, for instance a smart dressing applied to skin. The measurement settings will then need to be adjusted to keep the temperature increase below an acceptable value defined by the application's requirements. Reducing the measurement duty cycle appears to be an efficient way to mitigate heating to the desired level without substantially compromising performance, and can be used in conjunction with reducing the peak LED and MW powers.

\section{Conclusion}

We designed and experimentally realised a device termed ODMR-meter, which enables the contactless monitoring of ODMR in a target object, for instance a dressing or coating containing ODMR-active diamond particles for temperature monitoring applications. The central component of the apparatus is a handheld readout head which integrates all the necessary optical components and a microwave antenna, and was carefully designed to optimise device performance while providing a sufficient working distance for reading an object by hand. Our prototype also includes a control box connected to a laptop computer, making it an overall relatively compact and portable apparatus. Both readout head and control box are primarily made of commercially available, off-the-shelf components. We tested our prototype using a bulk diamond as the object and demonstrated a temperature sensitivity of $10~{\rm mK}/\sqrt{\rm Hz}$ in static conditions. In addition, we demonstrated the feasibility of temperature monitoring with handheld operation. The time for each temperature reading, which characterises the measurement speed, was 42 ms in our handheld demonstration, and can be reduced to $\sim1$~ms with simple hardware improvements. Likewise, we project that the sensitivity can be improved to below $\sim1~{\rm mK}/\sqrt{\rm Hz}$. However, the accuracy of the temperature measurement is currently poor with possible systematic errors of several K. Careful calibration of the temperature response and well parameterised fitting are expected to bring the accuracy to below 1~K. Finally, heating induced by the measurement is significant but can be efficiently mitigated by adjusting the measurement duty cycle. 

With the accuracy improved, the presented device may find immediate use in medical and industrial applications where accurate thermometry is required, via the use of smart dressings applied to skin or smart coatings applied to tools, for instance. A specific proposed application is the monitoring of infection of a wound without the need for dressing removal, which will require sub-K accuracy. Monitoring the temperature of drilling tools used for mining, while in operation, is another potential application. The device can be easily extended to temperature imaging by replacing the photodiode with a camera~\cite{Scholten2021}, which could open further applications. In both cases (single point measurement and imaging), the key advantage over infrared-based methods for non-contact thermometry is the superior accuracy, which often limits the reliability of infrared thermometers, although currently the ODMR-meter comes at a higher cost, size, weight, and power consumption (see Appendix~\ref{App:comparison} for a detailed comparison of the two technologies). 
Beyond temperature, similar smart coatings could be used to monitor the stray magnetic field from a magnetic object, for instance to determine its magnetization with high accuracy. For this magnetometry application, the presented ODMR-meter could be used without modification apart from a modified ODMR analysis procedure in order to extract the magnetic field value~\cite{Rondin2014}. Finally, we note that the device design can be easily adapted to accommodate other ODMR-active materials. For instance, the boron vacancy defect in hexagonal boron nitride has recently been shown to be capable of temperature sensing~\cite{Gottscholl2021,Healey2022}, and would only require a different emission filter.

\begin{acknowledgments}
This work was supported by the Australian Research Council (ARC) through grants FT200100073 and DP220100178, and by an NHMRC Ideas Grant (ID2002254). I.O.R. is supported by an Australian Government Research Training Program Scholarship. We thank Marco Capelli for providing the test diamond.
\end{acknowledgments}

\appendix

\section{Modelling details} \label{App:model}

\begin{figure}[htb!]
    \centering
    \includegraphics[width=8.5cm]{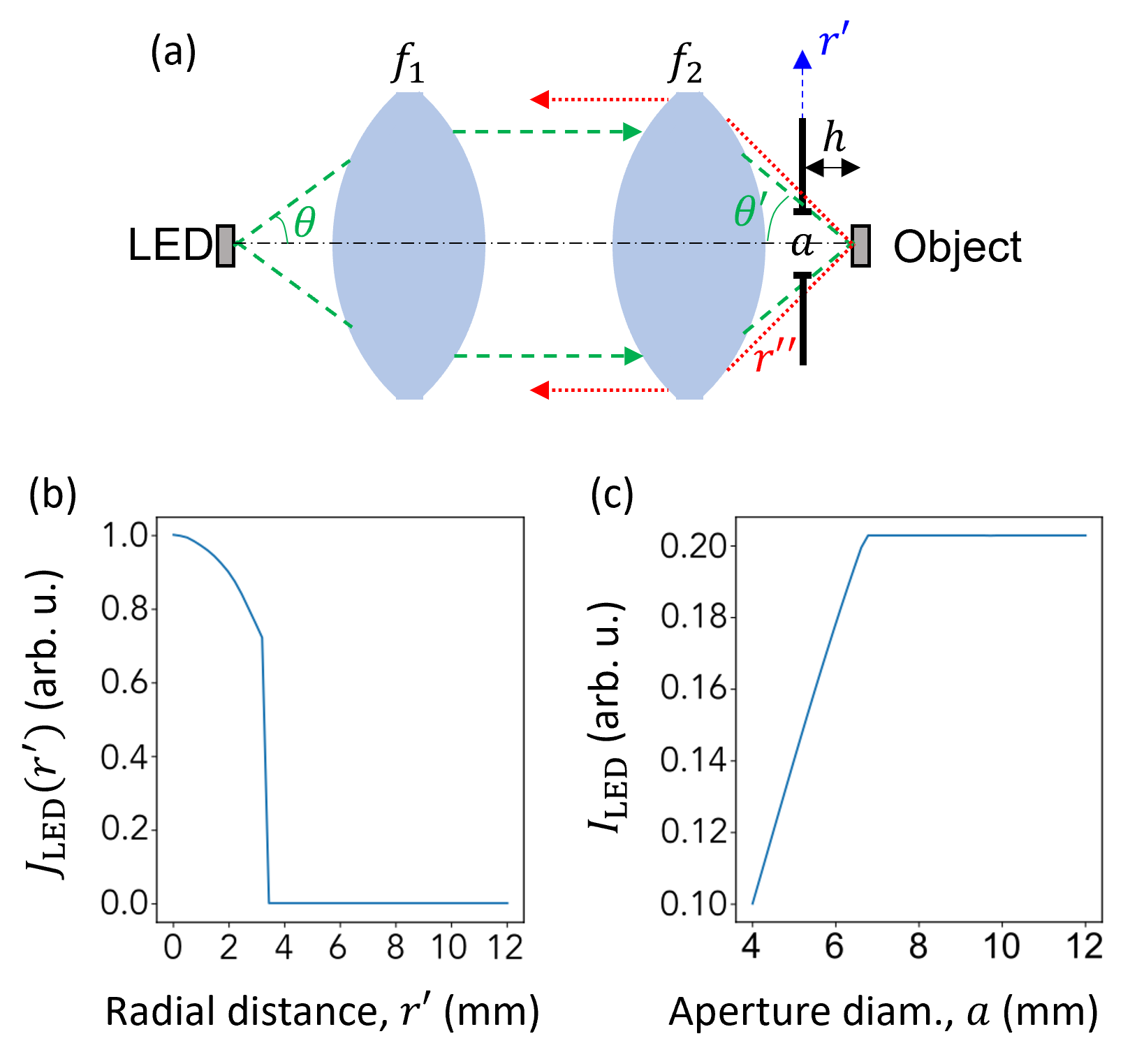}
\caption{(a) Model used to describe the effect of the MW loop (aperture $a$) on the LED light intensity received by the object and the amount of PL collected by the objective lens. (b) LED intensity as a function of the radial distance in the plane of the aperture. (c) Total LED power received by the object as a function of aperture diameter.}
    \label{fig:model}
\end{figure}

 Here we detail the model used to describe the data in Fig.~\ref{fig:design_tradeoffs}a. We treat the LED as a point source and use the angular intensity distribution $f(\theta)$ given by the supplier. The emitted LED light is collimated by a lens of focal length $f_1=20$\,mm [Fig.~\ref{fig:model}a]. The collimated LED light is then focused onto the object (diamond sensor, treated as a point) by a second lens of focal length $f_2=16$\,mm. The PL emitted by the object, assumed to be isotropic (i.e. intensity independent of emission angle), is collimated by the second lens and focused by a third lens onto the photodiode (not shown in the schematic). The PCB is modelled as an aperture of diameter $a$ (which in practice in slightly smaller than the MW loop diameter $d$) located a distance $h$ from the object. This aperture limits both the amount of total LED light reaching the object ($I_{\rm LED}$), and the amount of PL collected ($I_{\rm PL}$).
 
From simple ray tracing, the angular distribution $f(\theta)$ can be converted to an intensity in the plane of the aperture as a function of radial distance, $J_{\rm LED}(r')$, plotted in Fig.~\ref{fig:model}b, showing that the LED light is confined to $r'<3.4$\,mm. As a result, only aperture diameters $a<7$\,mm will clip the LED light. The total LED power ($I_{\rm LED}$) reaching the object (calculated at the focus point of lens 2) is calculated directly from $f(\theta)$:
    \begin{align}
    I_{\rm LED} &\propto \int_{\rm aperture} f(\theta(\theta')){\rm d}\Omega\\
    &\propto \int_0^{\arctan{\frac{a}{h}}}f\left(\frac{f_2}{f_1}\theta'\right)\sin\theta'{\rm d}\theta' \label{eq:LED}
    \end{align}
where ${\rm d}\Omega=\sin\theta {\rm d}\theta {\rm d}\phi$ is the differential solid angle. $I_{\rm LED}$ is plotted as a function of the aperture diameter in Fig.~\ref{fig:model}c, reaching a plateau for $a>7$\,mm as expected. Note, we use the symbol $I$ to refer to the power (in W, called intensity in the main text), and $J$ for the power density (in W/m$^2$). 

On the other hand, under the isotropic emission assumption the PL intensity in the plane of the aperture as a function of radial distance $r'$ is:
    \begin{align}
    J_{\rm PL}(r') &\propto \frac{I_0}{{r''}^2}
    \end{align}
where ${r''}^2=h^2 + {r'}^2$ is the distance from the object and $I_0$ is the emitted power. The total PL power through the aperture is then the intensity integrated over the solid area with constant radius ${r''}^2=R^2=a^2+h^2$:
    \begin{align}
    I_{\rm PL}&\propto\int_{\rm aperture} \frac{I_0}{R^2} {\rm d}A\\
    &\propto \int_0^{\arctan{a/h}} \frac{I_0}{R^2}R^2\sin\theta {\rm d}\theta\\
    &\propto I_0\times\left(1-\frac{1}{\sqrt{1+\frac{a^2}{h^2}}}\right) \label{eq:PL}
    \end{align}
where ${\rm d}A=R^2{\rm d}\Omega$ is the differential area. Eq.~\ref{eq:PL} is a saturation law function of $a/h$. Note, this differs from the quadratic scaling as a function of acceptance angle (or numerical aperture, NA) owing to the planar nature of the aperture, versus the spherical wavefront from the point source.

Combining Eq.~\ref{eq:PL} and \ref{eq:LED} and assuming the emitted power is proportional to the received power, i.e. $I_0\propto I_{\rm LED}$, we obtain 
    \begin{align}
    I_{\rm PL}&\propto\left(1-\frac{1}{\sqrt{1+\frac{a^2}{h^2}}}\right)\int_0^{\arctan{\frac{a}{h}}}f\left(\frac{f_2}{f_1}\theta'\right)\sin\theta'{\rm d}\theta' \label{eq:PLtot}
    \end{align}
This PL power is then transduced to a voltage by the photodiode, in a process independent of $a$. Thus, the solid line in Fig.~\ref{fig:design_tradeoffs}a is simply Eq.~\ref{eq:PLtot} with a pre-factor adjusted to best fit the experimental data. Note, in Fig.~\ref{fig:design_tradeoffs}a $I_{\rm PL}$ is plotted as a function of the MW loop diameter, which is $d=a+1.4\,{\rm mm}$ due to the finite width of the strip line forming the loop and manufacturing margins.

\begin{table*}[t!]
\begin{tabular}{|c|c|c|}
\hline
  & Fluke 62 MAX+ handheld  & ODMR-meter + bulk diamond  \\ 
  &  infrared laser thermometer~\cite{Fluke} & (this work) \\
\hline
Temperature range & $-30^\circ$C to $650^\circ$C & $-200^\circ$C to $300^\circ$C (expected~\cite{Chen2011,Toyli2012}) \\ 
\hline
Measurement rate & 2 Hz & 4 Hz (Fig.~\ref{fig:data}) \\
& (based on response time of 500\,ms) & 24 Hz (Fig.~\ref{fig:video}) \\
& & 0.4 Hz (Fig.~\ref{fig:heating}, at 10\% duty cycle) \\ 
\hline
Sensitivity & $350~{\rm mK}/\sqrt{\rm Hz}$ & $10~{\rm mK}/\sqrt{\rm Hz}$ (Fig.~\ref{fig:data})\\
& (based on repeatability of readings of $\pm0.5$\,K) & $30~{\rm mK}/\sqrt{\rm Hz}$ (Fig.~\ref{fig:heating}, at 10\% duty cycle) \\ 
\hline
Accuracy & $\pm1$\,K for $T>0^\circ$C (if emissivity is perfectly known) & $<1$\,K \\ 
 & $\pm2$\,K for $-10^\circ$C~$<T<0^\circ$C & (expected with parametrised fitting) \\ 
 & $\pm3$\,K for $-30^\circ$C~$<T<10^\circ$C &    \\ 
 \hline
Working distance & Variable: 12\,cm for a 1-cm object, 12\,m for a 1-m object & Fixed: 4\,mm \\
\hline
Additional requirement & Object's emissivity must be known & Object must be coated with diamond sensors \\
\hline
Size & $1.1$\,L & 0.9\,L (readout head) $+~10$\,L (control box)  \\ 
\hline
Weight & $255$\,g & $430$\,g (readout head) $+~5$\,kg (control box)  \\
\hline
Power consumption & 200\,mW & 20\,W  \\ 
\hline
Cost & \$200 & \$5,000  \\ 
\hline
\end{tabular}
\caption{Comparison of the performance and physical specifications of the ODMR-meter demonstrated in this work with a typical industry-grade infrared thermometer~\cite{Fluke}. For the ODMR-meter, the values for the size, weight, power consumption, and cost, are approximate and do not include the laptop nor the diamond sensor.}	
\label{tab:comparison}
\end{table*}

\section{Comparison with infrared thermometry} \label{App:comparison}

The main existing method for non-contact thermometry is infrared-based thermometry, which relies of the measurement of the thermal radiation emitted by the object. If the emissivity of the object's surface is known, its temperature can be determined. The main limitation of infrared thermometers is their relatively poor accuracy. Manufacturers typically quote an accuracy of $\pm1$\,K, but this assumes a perfectly known emissivity. In practice, variations in the emissivity caused by changing conditions (e.g., soiling of the surface or moisture), which are difficult to adjust for, can lead to additional errors of several K. 

Thus, the key expected advantage of an ODMR-based thermometer is the improved accuracy; as discussed above (Sec.~\ref{sec:accuracy}), an absolute accuracy better than 1\,K should be achievable using the presented device, independent of soiling or moisture. Nevertheless, infrared thermometers are a mature technology available at a lower cost and for a lower size, weight and power consumption than our current ODMR-meter, while also allowing a much larger working distance. A comparison between the ODMR-meter presented in this work and a typical industry-grade infrared thermometer (Fluke 62 MAX+ handheld infrared laser thermometer~\cite{Fluke}) is given in Table~\ref{tab:comparison}. With improved engineering following the path of compact NV magnetometers~\cite{Zhang2017,Kim2019,Sturner2019,Webb2019,Zheng2020,Huang2021,Sturner2021,Mariani2022,Wang2022,Xu2022}, we project that the ODMR-meter could become competitive in terms of cost, size, weight and power. We note that the ODMR-meter also features a better sensitivity (nearly $10\times$ better) than the infrared thermometer, making the ODMR technology overall promising for applications requiring non-contact temperature monitoring with high accuracy and sensitivity.

\bibliography{ODMeteR_bib}

\end{document}